\documentclass{article}
\usepackage{epsfig}

\begin{document}
\title{Experiences with iterated traffic microsimulations in Dallas}

\author{Kai Nagel\\
~\\
Los Alamos National Laboratory, TSA-DO/SA Mail Stop M997,\\
Los Alamos NM 87544, USA, {\tt kai@lanl.gov}}

\maketitle

\begin{abstract}
This paper reports experiences with iterated traffic microsimulations
in the context of a Dallas study.  ``Iterated microsimulations'' here
means that the information generated by a microsimulation is fed back
into the route planner so that the simulated individuals can adjust
their routes to circumvent congestion.  This paper gives an overview
over what has been done in the Dallas context to better understand the
relaxation process, and how to judge the robustness of the results.
\end{abstract}

\section{Introduction}
The advent of ever more powerful workstation computers makes large
scale transportation microsimulation projects feasible.  At the same
time, traditional tools are having a hard time treating all the
complexities of modern congested transportation systems.  As a result,
there has been considerable progress in the area of transportation
microsimulation in recent years, both in terms of theoretical
understanding of
micro-models~\cite{Nagel:flow:pre,Krauss,Schadschneider,Sugiyama} and
in terms of practical
implementations~\cite{Rickert:phd,Rickert:Nagel:DFW,INTEGRATION:overview,FVU-NRW,PARAMICS:Supercomp,Esser,Chopard}.
Yet, having good transportation microsimulations is only part of the
problem.  When running such a microsimulation, one finds oneself
confronted with the question of how to ``drive'' it: when a driver in
the microsimulation approaches an intersection, how does she know
which way to proceed?

The traditional answer to this question are turning percentages: for
example 50\% of the vehicles go straight, 20\% left, and 30\% right.
One is, though, immediately faced with a data collection problem: It
is improbable that one knows the turn counts for all intersections of
a regional area; and if one does not, one needs methods to
``generate'' the missing information~\cite{Esser}.  Also, for
transportation {\em planning\/} purposes one recognizes quickly that
these numbers are not very useful because they change easily with
infrastructure changes.

A somewhat better way is to use origin-destination matrices.  Yet, the
problems are the same: The information is not available, especially
for non-work trips; and these matrices change with major
infrastructure changes.

An answer to this is to start from demographics: It is improbable that
people's income changes or that they move in a matter of weeks in
response to transportation infrastructure changes.  From there, for a
{\em micro\/}simulation project, the first step is to generate
``synthetic'' populations from the demographic
data~\cite{Beckman:populations}.  The next task is to derive activities
(sleep, work, shop, $\ldots$) for each member of a synthetic
population, and then to derive the transportation demand from this.
So far no project has succeeded in completely executing this program,
i.e.\ to use activities based on demographics to ``drive'' a
transportation microsimulation.

There are considerable challenges ``on the way''.  For example, even
when given a time-dependent origin-destination matrix and a traffic
network, how does one allocate the trips to the network?  Traditional
assignment methods can be shown to be dynamically inconsistent under
heavily congested conditions.  One of the major problems is that for
links where demand is higher than capacity, the link travel time
depends on for how long the congested condition has been in place and
not just on demand only, and thus the traditional link travel time
functions are invalid.

A way out is to replace the traditional cost functions by a
microsimulations.  In short, one allocates traffic streams to the road
network, runs the microsimulation and collects link travel times,
re-allocates some of the traffic streams, re-runs the microsimulation,
etc.  This mimics ``day-to-day'' dynamics, i.e.\ ``over night'' some
drivers decide to try a new route the next day.  Obviously, one 
faces considerable challenges, such as: Which fraction of the
population should be re-planned?  How do we reach fast convergence of
the process?  Does the process converge at all?  Can we measure
convergence?  Is a real traffic system converged?  This paper will
report results related to these questions in the context of a Dallas
study.  After providing the context (Sec.~2) and showing the first
results (Sec.~3), the paper will summarize systematic feedback studies
(Sec.~4) and possible ``structural'' convergence criterions (Sec.~5).
After that, this paper will look at different aspects of the
``robustness'' of results; first under the change of a random seed
(Sec.~6), then under the change of the complete microsimulation
(Sec.~7), then in comparison to reality (Sec.~8).  Section~9 discusses
the robustness question in somewhat more general, pointing out that
one first needs to define the question that the simulation is supposed
to answer.  The paper is concluded by a short summary (Sec.~10).

\section{The context}

The context of the results reported here is the Dallas/Fort Worth case
study of the TRANSIMS project.  The goal of the case study was to
demonstrate that the approach has the general capability of generating
output that is useful for stake-holder analysis and scenario
evaluation~\cite{Beckman:etc:case:study}.  The setting of the study
was a 5~miles times 5~miles area (``study area'') around the busy
freeway intersection between the LBJ freeway and the Dallas North
Tollway north of Dallas downtown.  The problem was approached by
starting out with a ``focused network'' and a production-attraction
(PA) matrix.  The focused road network contained {\em all\/} streets
inside the study area, but got considerably thinner with increasing
distance from the study area.  A production-attraction matrix is
essentially a 24-hour origin-destination matrix with some land-use
information included (trips from work to home are entered as trips
from home to work, with the result that the zones where trips
originate in these matrices need to be residential zones).  The PA
matrix for Dallas/Fort Worth contained approximately 10~million trips
during a 24-h period.

Since TRANSIMS is a microscopic approach, the first thing that was
done was to decompose the PA matrix into individual trips.  This was
done using a 24-hour time use function, i.e.\ each trip had a starting
time, and the distribution of starting times reflected rush period
traffic.  Out of these trips, only the trips starting between 5am and
10am were considered.  These trips were routed along the road network,
and only the trips which went through the study area were retained
(about 300\,000 trips).  This set of trips, sometimes called ``initial
planset'', forms the basis of all studies presented here.
See~\cite{Beckman:etc:case:study,Nagel:Barrett:feedback} for further
details.

\section{First results}

\begin{figure}[t]
\epsfxsize\hsize
\epsfbox[36 129 576 664]{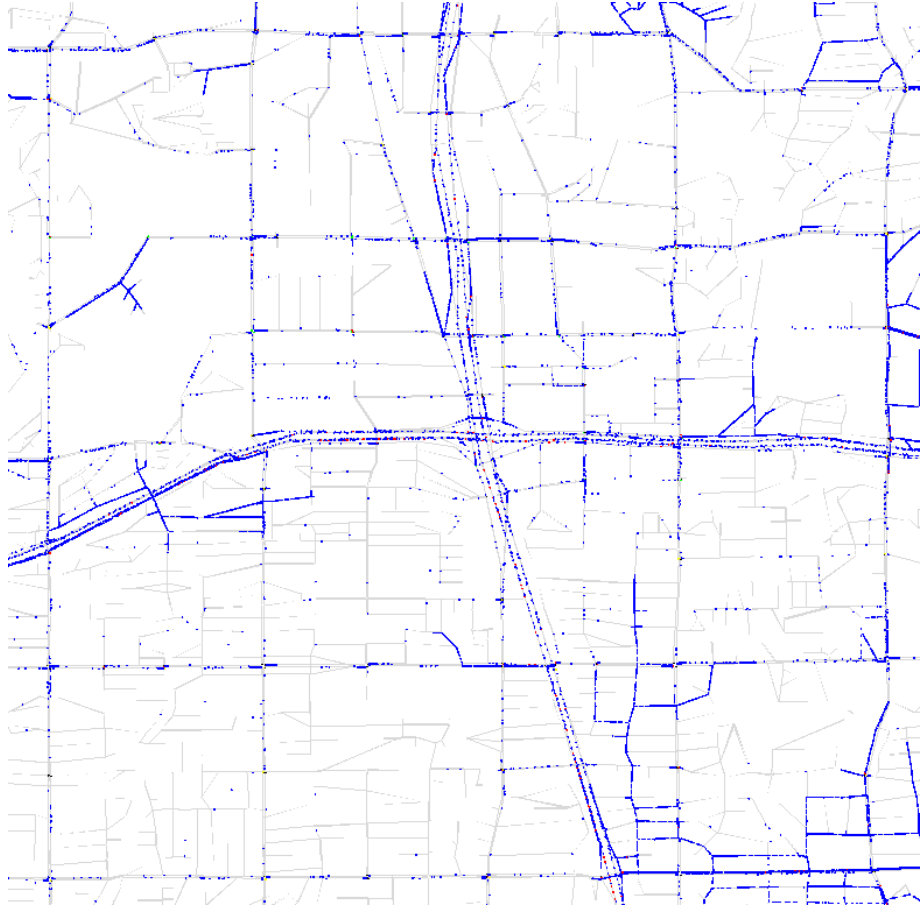}
\caption{\label{0-it-9:30}%
Microsimulation run based on initial planset, snapshot at 9:30~am.
}
\end{figure}

Just using the routes from the initial planset and sending them
through a traffic simulation\footnote{Note that this remark implies
that we are using microsimulations based on {\em route plans}, i.e.\
each driver knows before her trip starts exactly the sequence of links
she wishes to take through the network.} typically generates a result
as in Fig.~\ref{0-it-9:30}, where many streets are occupied by jammed
traffic.  The problem behind this is, obviously, that the route
generation phase for the initial planset did not take into account
what other people were going to do.  Yet, predicting what everybody
else will do in a straightforward way is impossible; imagine yourself
in a situation where you wake up one morning in a city where you need
to go to work, and the only two pieces of information you have is a
map of the transportation system and the information that everybody
else is in the same situation as you.

\begin{figure}[t]
\epsfxsize\hsize
\epsfbox[36 113 578 664]{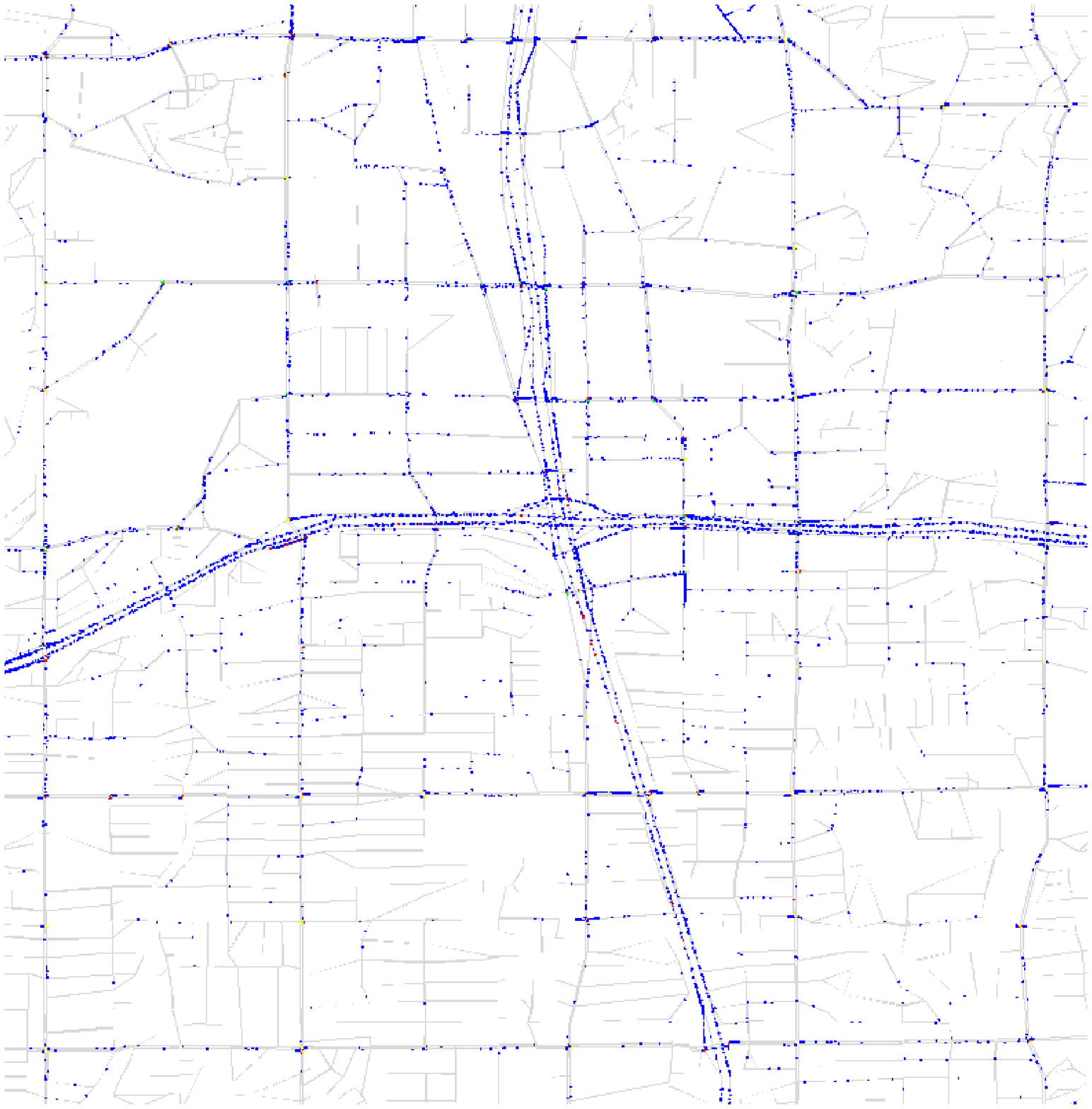}
\caption{\label{14b-it-9:30}%
Microsimulation run based on the planset after 14~iterations, snapshot
at 9:30~am.  Compare to Fig.~\protect\ref{0-it-9:30}.
}
\end{figure}

The way we solve this is by iterations between microsimulation and
planner.  A possible method is the following: the microsimulation is
run, link travel times are extracted from the microsimulation, a
certain percentage of the travelers computes new routes based on these
link travel times, the microsimulation is run again with these new
plans, etc.  New routes are computed using time-dependent fastest path
based on the starting time, starting location, destination, and link
travel times from the microsimulation.  Other iteration schemes are
possible, see, e.g.,~\cite{Nagel:NRW}.

\begin{figure}[t]
\epsfxsize\hsize
\epsfbox[36 141 576 652]{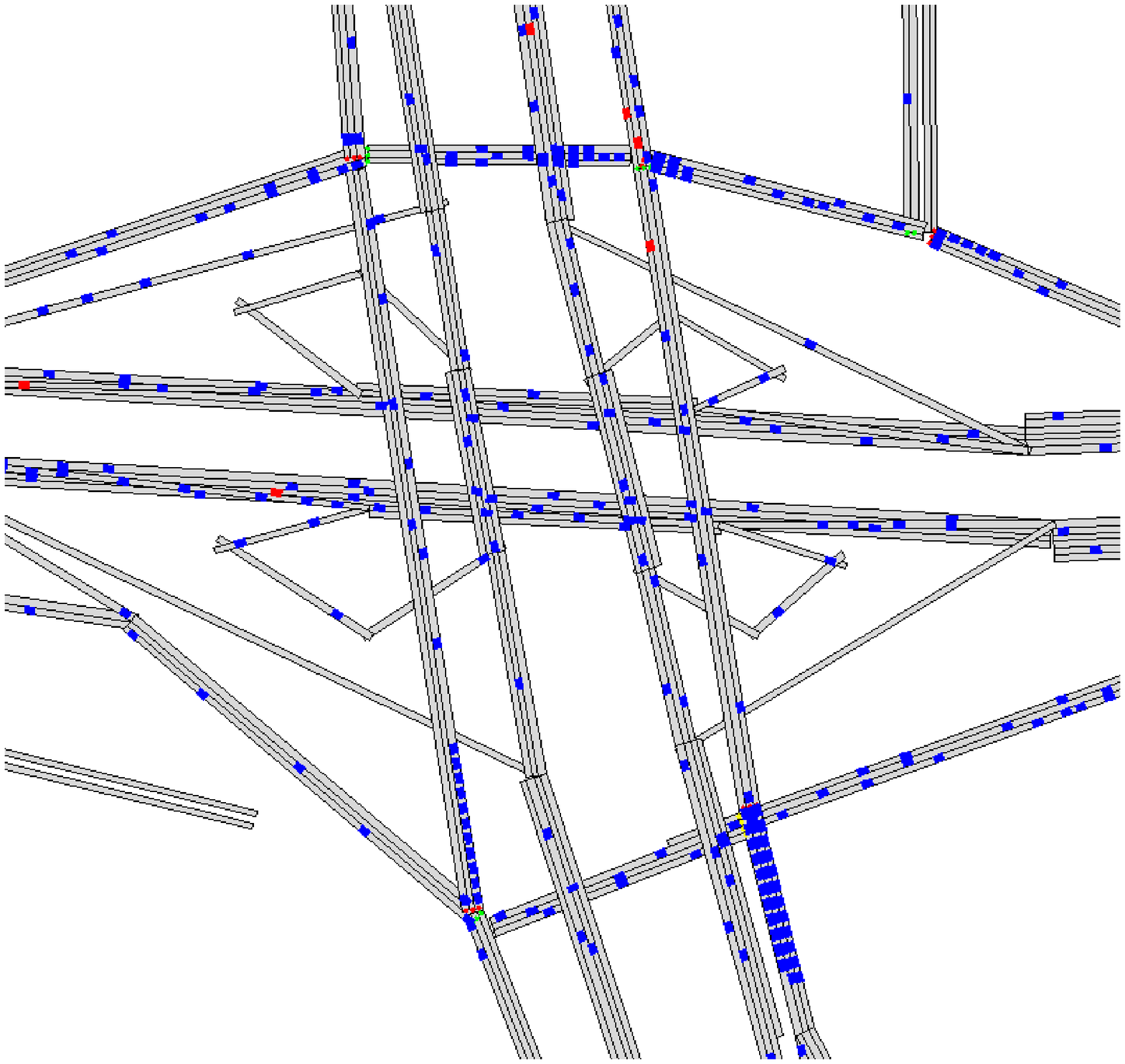}
\caption{\label{14b-it-9:30-detail}%
Microsimulation run based on the planset after 14~iterations, snapshot
at 9:30~am, detail.  Traffic lights are shown by their colors at the
end of links.  Cars can be color coded according to specified criteria
-- in this specific example, cars that were not able to follow their
intended plans because they could not get into the desired lane are
shown red.
}
\end{figure}

An open question here is which percentage of the travelers should be
re-planned.  First, we started out with a fairly high percentage,
10\%, and used that until the results started oscillating, which was
after the 7th iteration.\footnote{The reason for the oscillation is
easy to understand: Assume you have two route alternatives, and one is
slightly faster.  If you re-route a certain percentage of people, than
the other alternative will be faster.  Without additional measures,
this will be an undamped oscillation.}  In iterations number 8--12,
5\% of the trips were re-routed; in iterations number 13--14, 2\% of
the trips.  A typical result after 14~iterations can be seen in
Figs.~\ref{14b-it-9:30} and~\ref{14b-it-9:30-detail}.  Clearly, the
traffic jams have cleared, and traffic is not only more evenly
distributed, but because the system is more efficient, traffic does
not back up.  Fig.~\ref{14b-it-9:30} looks ``plausible'', whereas
Fig.~\ref{0-it-9:30} does not.  Further information can be found
in~\cite{Nagel:Barrett:feedback}.

\section{Systematic feedback studies}
\label{feedback:studies}

Using a relaxation scheme such as the above is not very practical
(because it needs human supervision) and scientifically not very
convincing: Questions such as ``Does the process converge?  If so, can
we quantify `distance' from the converged state?  Is real traffic
converged?''  come up.  One next step is thus to more systematically
test relaxation schemes and relaxation properties.  (Since we do not
know if the process ``converges'' in the mathematical sense, we will
talk about ``relaxation'' instead.)
Rickert~\cite{Rickert:phd,Rickert:feedback} has run such tests in the
same Dallas/Fort Worth context, but with a different
microsimulation~\cite{Rickert:Nagel:DFW}.  His microsimulation is
somewhat less realistic (most important points: no signal plans, no
turn pockets, no lane changing for turning behavior), but it currently
runs more than 20~times faster than the microsimulation used in the
last section.  Some comparisons between the results obtained by both
microsimulations will be shown below; we believe using a different
microsimulation will not have any significant impact on the general
results that will be discussed in this section.

\begin{figure}[t]
\epsfxsize0.8\hsize
\centerline{\epsfbox{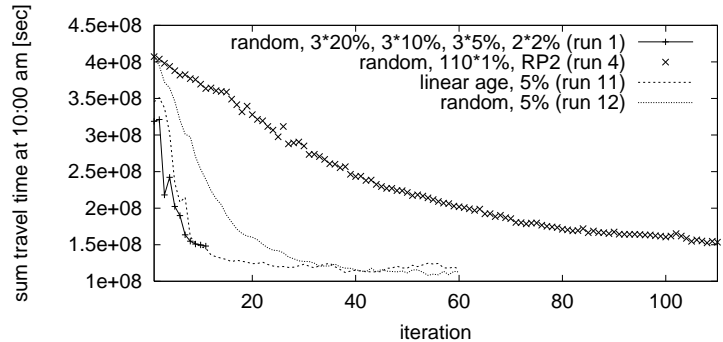}}
\caption{\label{Marcus:by:iteration}%
Sum of all travel times as a function of the iteration.  Clearly, the
sum of all travel times decreases with the iterations because
congestion clears and the efficiency of the whole system increases.
Different iteration schemes result in different relaxation speeds.
From~\protect\cite{Rickert:feedback}.
}
\end{figure}

Fig.~\ref{Marcus:by:iteration} shows the sum of all travel times as a
function of the iteration number for different iteration schemes.  For
example, the full line (also marked by $+$) is for the same relaxation
scheme as described in the last section.  The data marked by the
$\times$ symbol is for an iteration series where in each iteration
only 1\% of the travelers was re-planned.  The dotted lines are for
iteration series where 5\% of the travelers were re-planned; in one
case these 5\% were selected randomly, in the other case they were
selected according to ``age'': A traveler who had not tried a new
route for a long time was more prone to re-planning.  Several
observations can be made from this plot:\begin{itemize}

\item
The simulations relax to a value of approx.\ $1.1 \cdot 10^8$~seconds
for the sum of all travel times.  Clearly, the iteration described by
the full line and ``$+$'' is not yet there.

\item
The iterations marked by the $\times$ symbol are relaxing very slowly.
The reason for this is that 1\% re-planning means that the probability
of {\em not\/} having been re-planned after 110~iterations is 
$0.99^{110} \approx 0.33$, i.e.\ about one third of all trips still
follow their initial route which has been computed without {\em any\/}
feedback information.

\item
As the line for ``age-dependent'' re-planning shows, much faster
relaxation schemes are possible even when they are non supervised.

\end{itemize}
For further information, see~\cite{Rickert:phd,Rickert:feedback}.

\section{Characteristics of fastest paths in relaxed vs.\ unrelaxed
traffic situations}

The above quantity, sum of all travel times, has the disadvantage that
as a relaxation criterion it is only meaningful in the context of a
relaxation series: One can follow its behavior, and from that one can
decide that the series is now ``relaxed'' or not.  This approach is
not useful for answering the question where the ``real'' system is.
Is it somewhere along such a relaxation line, but not quite at the
bottom?  Is it totally different?  The sum of all trip times is a
number that cannot be compared across different networks: That number
would, for example, depend on the size of the region.

For answering such a question, one needs a more ``structural''
approach, one that looks ``directly'' at the properties that
supposedly relax.  The property that relaxes in the above approach is
the ``incentive to deviate''.  Our behavioral assumption is that
people, given the starting time and location and the destination, will
switch routes until they find a reasonably fast route.  To a certain
extent, this follows the traditional assumption both for the
Wardropian equilibrium in transportation and for a Nash equilibrium in
game theory in that rational players choose their best strategy ($=$
fastest route), based on the assumption that everybody else is
rational.  Yet, in reality people do not search for the fastest route
at all costs; also, the above relaxation scheme when inspected closer
reveals that in it people do not select the route with the fastest
{\em expected\/} travel time, even not in the average.  

So the question becomes if we can measure a quantity which reflects
the incentive to deviate; and if so, if this quantity decreased with
the iteration numbers.  A different way to formulate the question is:
How much additional time would you need for your second-best route?
One of the problems here is that ``second-best'' route is not
well-defined; simply computing $k$--shortest paths in an algorithmic
way will generate many solutions that most people will not consider
alternatives; such as leaving a freeway at an off-ramp, crossing the
intersection, and getting on again at the other side of the
intersection.  Yet, other approaches based on ``reasonableness''
cannot be considered satisfying for the present
question~\cite{Rilett:reasonable:paths}.

\begin{figure}[t]

\epsfxsize0.8\hsize
\centerline{\epsfbox{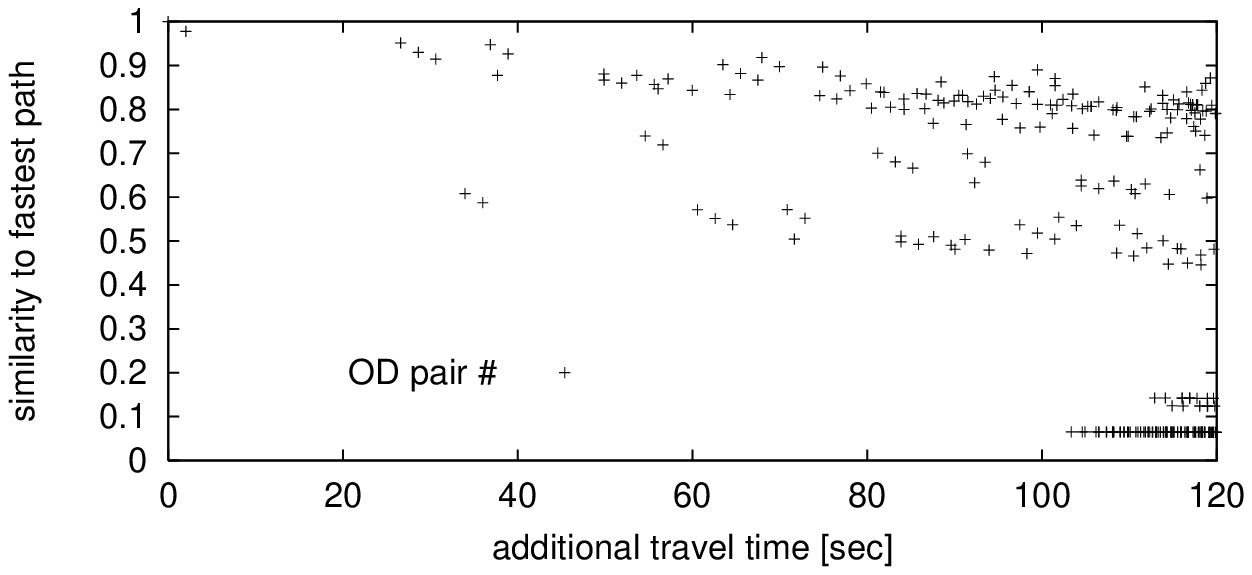}}

\epsfxsize0.8\hsize
\centerline{\epsfbox{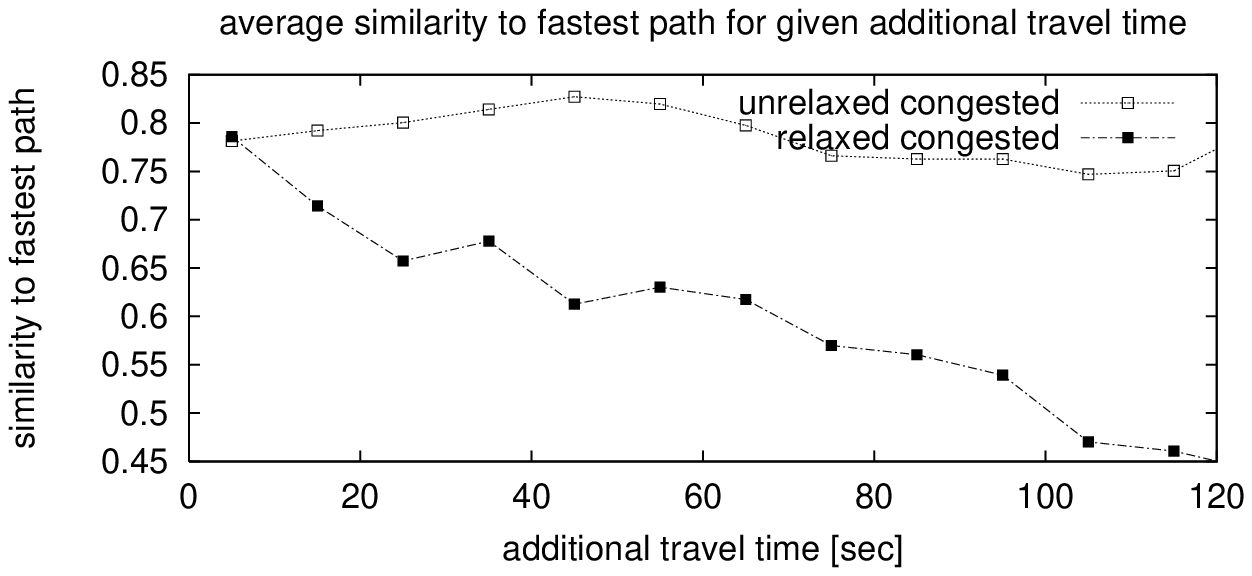}}

\caption{\label{ksp}%
Plotting ``similarity to fastest path'' as a function of ``additional
travel time''.  {\em Top:} Example.  One observes that, if one allows
more and more additional travel time, one obtains more and more
options that are very different from the fastest option.  {\em
Bottom:} Average over 955 origin destination pairs.  One sees that,
for given additional travel time, one has more diverse options in the
relaxed congested case.  From~\protect\cite{Kelly:Nagel:ksp}.
}
\end{figure}

In order to have a quantitatively sound criterion, we used
$k$--fastest paths but then calculated the ``difference to the fastest
path''.  By this, one can distinguish between routes which differ only
little from the fastest path and routes which differ a lot.  We then
plot that quantity as a function of, say, additional travel time.  The
information provided thus is (Fig.~\ref{ksp}a): Given I accept
$x$~seconds more travel time, how different are my possible routes
from my fastest route?  One can now calculate this information for
many different origin-destination pairs.  Averaging the resulting
information for given amounts of additional travel time results in
Fig.~\ref{ksp}b, i.e.\ the figure provides an answer to the question:
How different will my route be {\em in the average\/} from the fastest
route if I accept $x$~seconds of additional travel time?

Now, note that the two curves in Fig.~\ref{ksp}b have been computed
with two different link travel time sets: (i)~``unrelaxed congested''
uses link travel times obtained from an ``unrelaxed'' simulation
(i.e.\ based on the ``initial planset'') at a congested time of the
day; (ii)~``relaxed congested'' uses link travel times obtained from a
``relaxed'' simulation (i.e.\ after many iterations) at the same
congested time of the day.  Clearly, when accepting a certain amount
of additional travel time, the options one has under relaxed
conditions are in the average much more diverse than under unrelaxed
conditions.  In other words, many different routes will provide
near-optimum performance for the individual driver; the optimum
becomes ``flat''.  This means that the system arranges itself in a way
that the ``incentive to deviate'' is fairly small, because even if you
do not use the optimal route, the gains from switching are fairly
small.  For further information, see~\cite{Kelly:Nagel:ksp}.

\section{Random fluctuations}

\begin{figure}[t]
\epsfxsize\hsize
\epsfbox[37 135 575 657]{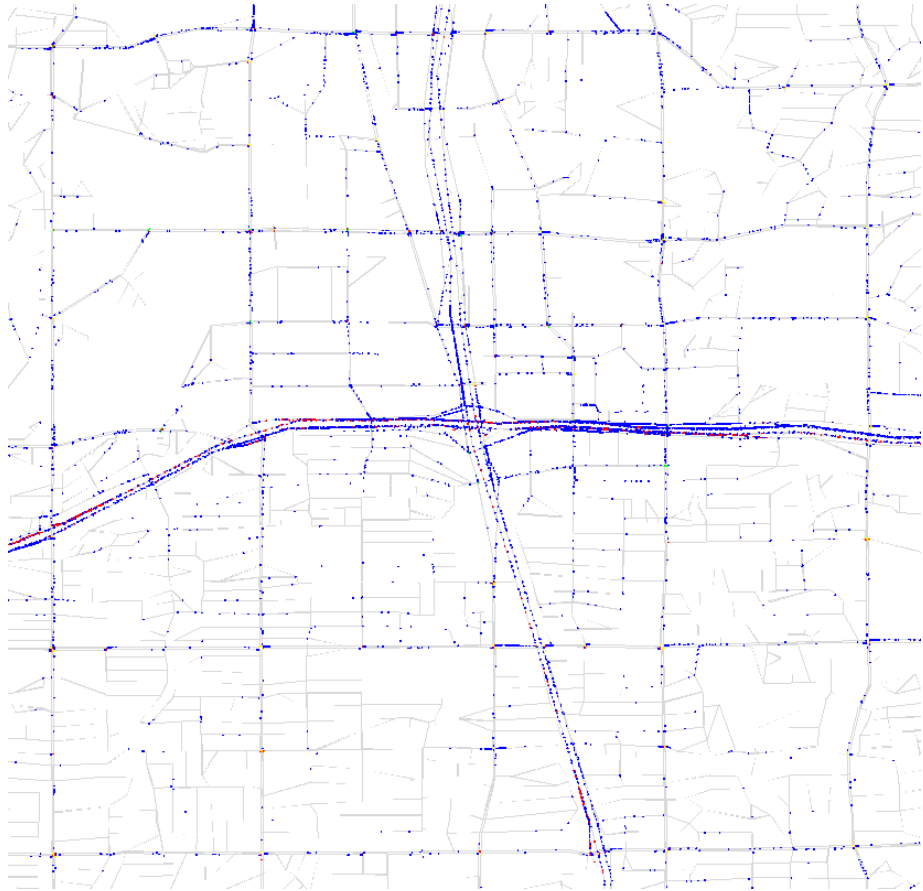}
\caption{\label{14-it-9:30}%
Demonstration of random fluctuations.  Microsimulation run based on
planset after 14~iterations, snapshot at 9:30~am.  Compare to
Fig.~\protect\ref{14b-it-9:30}, the only difference between the
runs is a changed random seed.
}
\end{figure}

Our microsimulations are stochastic, i.e.\ they use random numbers
during the computation of their dynamics.  That means that a simple
change of a random seed can change the entire dynamic trajectory of
the simulation.  Fig.~\ref{14-it-9:30} is an example how dramatically
the situationcan change with only the change of a random seed shown
(compare to Fig.~\ref{14b-it-9:30}).  Clearly, using stochastic
dynamics makes formal definitions, analytical proofs, and
computational proofs of convergence much harder.  Yet, we believe that
it is a necessary feature of reality; also, it makes the computational
search process more robust against getting ``stuck'' in implausible
situations.

A result of stochasticity is that transportation simulation projects
can, for any given question, at best return a ``distribution'' of
possible answers, i.e.\ deterministic answers are not possible.  Note
that, if the distribution is not Gaussian, then using the arithmetic
mean as a replacement for the deterministic answer can be misleading,
for example in the case of a bi-modal distribution.

This indicates that more systematic studies of the effects of
stochastic dynamics need to be done.

\section{Different microsimulations}

\begin{figure}[t]
\epsfxsize\hsize
\epsfbox[35 132 576 661]{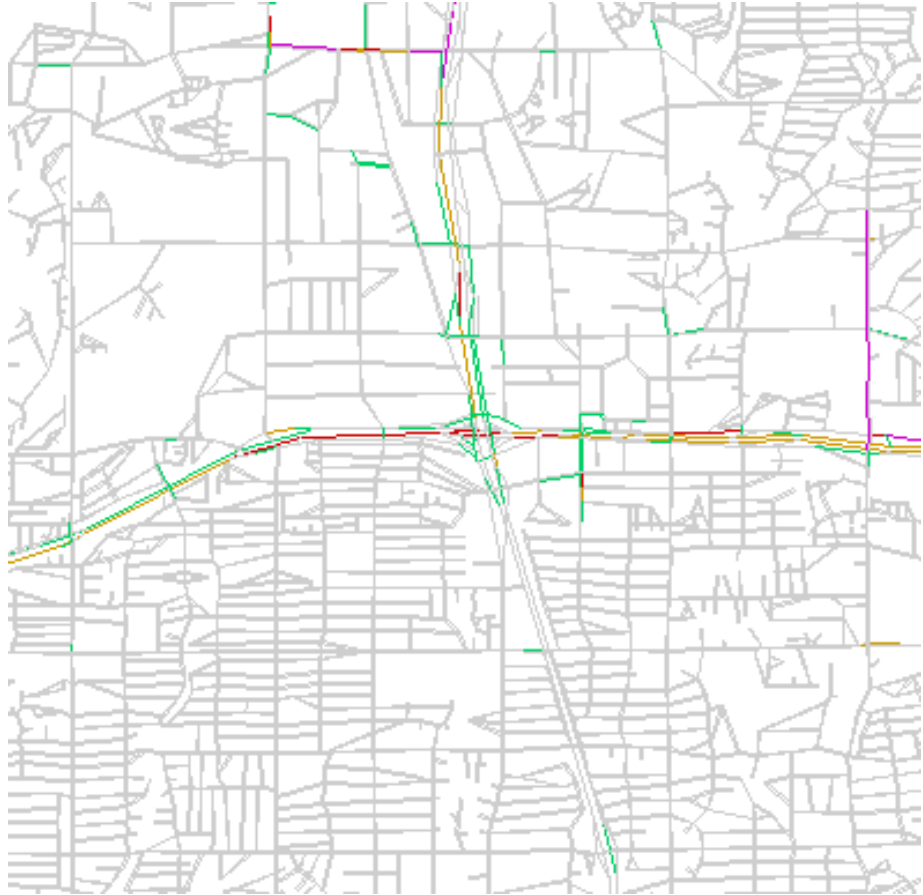}
\caption{\label{Marcus-9:30}%
Using a different microsimulation in the same study context.  Compare
to Figs.~\protect\ref{14b-it-9:30} and~\protect\ref{14-it-9:30}.
Unfortunately, a plot showing individual vehicles as in the other
plots was not available; but to a certian extent, a comparison is
possible.  Gray means link density (occupancy) below 10\%, green
means below 30\% (which corresponds roughly to capacity), yellow means
below 50\%, red means below 70\%, and purple means above
70\%. From~\protect\cite{Rickert:feedback}.
}
\end{figure}

At the current stage, it is unclear in how far transportation
simulation projects will be able to provide ``robust'' answers to
questions that are relevant to society.  As a minimum requirement, one
should be able to extract ``similar'' answers from different codes.
Fig.~\ref{Marcus-9:30} shows an example of using a different
microsimulation in the same context; in fact, the microsimulation is
the same as the one used for the results described in
Sec.~\ref{feedback:studies}.  The figure should be compared with
Figs.~\ref{14b-it-9:30} and~\ref{14-it-9:30}.  Clearly, there are
similarities.  Traffic on the eastern part of the east-west freeway is
in general heavy whereas most of the other simulation area turns out
to be outside the congested regime.  In some sense,
Fig.~\ref{Marcus-9:30} is in between Figs.~\ref{14b-it-9:30}
and~\ref{14-it-9:30}, which only differ by the initial random seed.
Beyond these somewhat general observations, comparison between
different microsimulations is not very well defined problem; see
Sec.~\ref{how} for a further discussion of this subject
and~\cite{Nagel:etc:comparisons} for more information.

\section{Comparison to reality}

In the context of the Dallas study, we had some turn counts from
selected intersections available.  Before turning to the results, some
details about the comparison have to be noted, which are a consequence
of the fact that the north-south freeway in the study area did not
exist in its full length before 1990:\begin{itemize}

\item
The main inputs for our microsimulations are the network and the trip
tables (PA matrices). 

\item
The trip table that we use is from before 1990, i.e.\ {\em before the
northern part of the north-south freeway existed}.

\item
The network that we use is from after 1990, i.e.\ {\em after\/} the
northern part of the north-south freeway was opened.

\item
The reality counts are from 1996, i.e.\ from a time where the network
was similar to the one in the simulation, but the trips had
adjusted to the existence of a much faster way to travel north-south
in the northern part of the study area.

\end{itemize}
In consequence, we would expect that our studies underestimate
north-south traffic in our study area compared to the 1996 counts.  

\begin{figure}
\epsfxsize\hsize
\epsfbox[36 126 575 665]{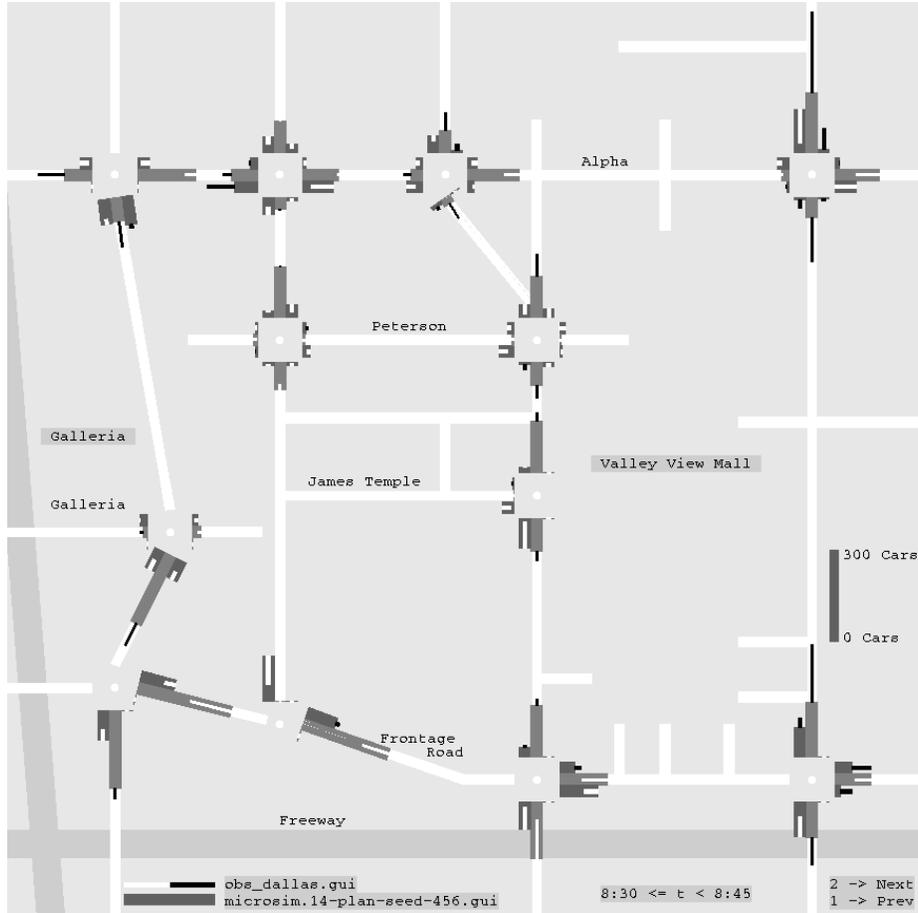}
\caption{\label{reality}%
Comparison with reality between 8:30am and 8:45am.  At most
intersections, there are three principal bars, one for right turns,
one for traffic going straight, and one for left turns.  For each of
these principal bars, the middle (black/white) bar shows the observed
value, the outside (gray) bar the value from the simulation.  For
example, for the intersection in the right upper corner, the
simulation is underestimating both southbound and northbound through
traffic. -- When interpreting this figure, one needs to know that the
rightmost north-south road and the topmost east-west roads are major
arterials, the dark grey roads are freeways, roads immediately
parallel to the freeways are frontage roads, and all other roads are
minor.  Only then it becomes clear that we are overestimating traffic
that goes through ``inconvenient'' routes, i.e.\ through sharp turns
and over minor roads.  Also, we underestimate north-south traffic in
general due to the data inconsistency described in the text.  Note
that we have made no attempt to ``calibrate'' the simulation to these
values.  From~\protect\cite{Nagel:etc:comparisons}.
}
\end{figure}

When looking at Fig.~\ref{reality}, this is clearly the case.  But
this is not the only feature.  One also notices that in fairly general
our simulation over-estimates traffic on minor roads and through
turns; this is most probably a consequence of the way the re-planner
works because it {\em only\/} looks at travel times and not, for
example, at inconveniences caused by sharp turns or by stop signs.
These results reflect work in progress; further results will be
published elsewhere~\cite{Nagel:etc:comparisons}.


\section{Robustness of (micro-)simulation results}
\label{how}

The previous two sections ask the question of how to compare different
microsimulations or of how to compare microsimulations with reality.
In principle, this should be easy: microsimulations generate
microscopic observables, and so we simply extract the same information
from the microsimulations and from reality and we compare them.  Yet,
it is unclear {\em which\/} information to exactly to extract: Which
is the most meaningful information?  For example, does the fact that
microsimulation XYZ has 20\% too many right turns on a certain
intersection really matter?  Or what is more important: 100~vehicles
more at the end of an already large traffic jam, or 100~vehicles more
on uncongested roads?

One needs to relate this problem to the question one attempts to
answer with a specific simulation project.  The level of fidelity of a
simulation project will always depend on the level of effort spent;
and it seems reasonable to us to adjust the level of effort to what is
really necessary for a given question.  Also, with a given question at
hand, the problem of comparing simulations becomes better defined: One
does not need any more to decide if two simulations are ``equal'', but
only if two different simulations give the same answer to the
specified questions.  This latter approach is much better accessible
to the statistical tools at hand; admittedly, it is not very
satisfying because it means that for any new question one needs, in
some sense, to start over.

\begin{figure}[t]
\epsfxsize0.8\hsize
\centerline{\epsfbox{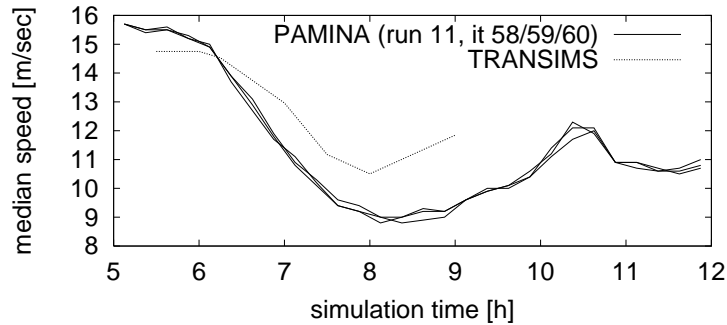}}
\caption{\label{comparison:ttimes}%
Comparison of median ``geographical'' speeds between two simulations.
What we mean by geographical speed is ``geographical distance''
divided by ``travel time'', i.e.\ {\em not\/} the average driving
speed.  This measures accessability of an area; in this case
accessabiliby of the interior of the simulation area when coming from
the outside.  From~\cite{Nagel:etc:comparisons}.
}
\end{figure}

Fig.~\ref{comparison:ttimes} presents one such possible result.
Plotted is the average speed for reaching the center of the simulation
area when coming from the outside.  The speed here is calculated by
``geographical distance divided by trip time'', i.e.\ it does {\em
not\/} reflect driving speed but is a measure of ``accessibility'':
Low speeds reflect that the destination is difficult to reach.  Our
plot is a simplified version of a type of questions that are really
important in the context of ``stake-holder analysis''.  For example,
does the introduction of a light rail make travel to the downtown area
faster for people who do not own a car? (Probably yes.)  Does it make
the same travel faster for people who own a car?  (Probably not, at
least not during uncongested conditions.)

Now note that Fig.~\ref{comparison:ttimes} only gives part of
the full answer.  We certainly see that in both simulations,
accessibility of the center of the simulated area drops during the
morning rush period.  We also see that the drop in accessibility is
different between both microsimulations.  Yet, if the question were
for example to find the time of worst accessibility during the morning
rush period, both simulations would give the same answer.

Last, but not least, note that this paper deliberately side-steps the
question of the trustworthiness of the microsimulation itself.  In
this paper, we have concentrated on ``macroscopic'' aspects, such as
large scale fluctuations and the interplay between route planner and
microsimulation.  Yet, it is also useful to test and document
microscopic aspects of traffic models.  The discussion of what to
measure here and how is ongoing; but there seems to be some slowly
growing consensus that certain building blocks of the flow
characteristics, such as flow through a stop sign as function of the
traffic on the major road, unrealistic as these examples may be,
should be part of these systematic tests.  The development of more
complex test suites would certainly be desirable.  For further
information on this in the context of TRANSIMS,
see~\cite{Nagel:etc:flow-char}.

\section{Summary}

Iterated transportation microsimulations provide a very powerful
addition to the tools of transportation planning.  Their power lies
both in the capability to represent time-dependent scenarios in their
``true'' time-dependent dynamics and in the possibility to access
individual, microscopic quantities directly.  Yet, one needs to note
that research in this area is at its beginning; no consistent theory
is available to help and even the building of intuition is in its
initial stages.  This paper provides a summary of what has been done
in the context of a TRANSIMS study of the Dallas area in order to
provide exactly some of this intuition.  These results can be
summarized as follows: (i)~Iterations between router and
microsimulation adjust routes in a way that traffic becomes
``plausible''.  (ii)~The number of iterations that are necessary until
the process is plausibly ``relaxed'' can be significantly reduced by
using ``intelligent'' relaxation schemes.  (iii)~It seems possible to
distinguish ``unrelaxed'' from ``relaxed'' conditions by looking at
fastest paths in both situations.  The result is interpretable in the
sense that in relaxed transportation systems, there is not much
difference in travel time between the strictly fastest path and very
different routes.  (iv)~In stochastic microsimulations, simply
changing the random seed can generate large fluctuations.  (v)~Using a
different microsimulation in the same scenario produces results that
look similar, but that are difficult to compare in general.
(vi)~Comparing to reality has the same caveats, plus the problem of
data consistency.  Nevertheless, the comparison provides useful
insights.  (vii)~In general, comparison between microsimulations need
to be geared to specific questions.  An example of such a question is
provided.

\section*{Acknowledgments}

Los Alamos National Laboratory is operated by the University of
California for the U.S.\ Department of Energy under contract
W-7405-ENG-36.  This article is work performed under the auspices of
the U.S.\ Department of Energy.


\begin{thebibliography}{10}

\bibitem{Nagel:flow:pre}
K.~Nagel.
\newblock Particle hopping models and traffic flow theory.
\newblock {\em Phys. Rev. E}, 53(5):4655, 1996.

\bibitem{Krauss}
S.~Krauss~et al, this volume.

\bibitem{Schadschneider}
A.~Schadschneider~et al, this volume.

\bibitem{Sugiyama}
Y.~Sugiyama~et al, this volume.

\bibitem{Rickert:phd}
M.~Rickert, in preparation.

\bibitem{Rickert:Nagel:DFW}
M.~Rickert and K.~Nagel.
\newblock Experiences with a simplified microsimulation for the {Dallas/Fort
  Worth} area.
\newblock {\em International Journal of Modern Physics C}, 8(3):483--504, 1997.

\bibitem{INTEGRATION:overview}
M.~Van~Aerde, B.~Hellinga, M.~Baker, and H.~Rakha.
\newblock {INTEGRATION}: An overview of traffic simulation features.
\newblock {\em Transportation Research Records}, in press.

\bibitem{FVU-NRW}
Forschungsverbund f{\"u}r Verkehr und Umwelt (FVU) NRW. See
  {http://www.zpr.uni-koeln.de/Forschungsverbund-Verkehr-NRW/}.

\bibitem{PARAMICS:Supercomp}
G.D.B. Cameron and C.I.D. Duncan.
\newblock {PARAMICS}--{P}arallel microscopic simulation of road traffic.
\newblock {\em J. Supercomputing}, 10(1):25, 1996.

\bibitem{Esser}
J.~Esser~et al, this volume.

\bibitem{Chopard}
B.~Chopard, this volume.

\bibitem{Beckman:populations}
R.J. Beckman, K.A. Baggerly, and M.D. McKay.
\newblock Creating synthetic baseline populations.
\newblock {\em Transportation Research A, Policy and Practice},
  30A(6):415--429, 1996.

\bibitem{Beckman:etc:case:study}
R.J. {Beckman et al}.
\newblock {TRANSIMS} {D}allas/{F}ort {W}orth case study report.
\newblock {Los Alamos Unclassified Report LA-UR} to be released, Los Alamos
  National Laboratory, TSA-Division, Los Alamos NM 87545, USA, 1997.

\bibitem{Nagel:Barrett:feedback}
K.~Nagel and C.L.Barrett.
\newblock Using microsimulation feedback for trip adaptation for realistic
  traffic in {Dallas}.
\newblock {\em International Journal of Modern Physics C}, 8(3):505--526, 1997.

\bibitem{Nagel:NRW}
K.~Nagel.
\newblock Individual adaption in a path-based simulation of the freeway network
  of {Northrhine-Westfalia}.
\newblock {\em International Journal of Modern Physics C}, 7(6):883, 1996.

\bibitem{Rickert:feedback}
M.~Rickert~et al, in preparation.

\bibitem{Rilett:reasonable:paths}
D.~Park and L.R. Rilett.
\newblock Identifying multiple and reasonable paths in transportation networks:
  A heuristic approach.
\newblock {\em Transportation Research Record}, In press.

\bibitem{Kelly:Nagel:ksp}
T.~Kelly and K.~Nagel, submitted.
\newblock Also LA-UR 97-4453.

\bibitem{Nagel:etc:comparisons}
K.~Nagel~et al, in preparation.

\bibitem{Nagel:etc:flow-char}
K.~Nagel, P.~Stretz, M.~Pieck, S.~Leckey, R.~Donnelly, and C.L. Barrett.
\newblock {TRANSIMS} traffic flow characteristics, submitted.
\newblock Also LA-UR 97-3530.

\end{thebibliography}

\end{document}